\documentstyle[12pt]{article}
\author{Walter Simon\thanks{Supported by a fellowship from R\'egion
Centre. Present address: Institut f\"ur theoretische Physik der
Universit\"at Wien, Boltzmanngasse 5, A-1090 Wien, Austria. 
e-mail:  simon@ap.univie.ac.at}\\
D\'epartement de Math\'ematiques\\Facult\'e des Sciences\\Parc de Grandmont\\
F-37200 Tours, France}
\title{Conformal positive mass theorems}
\date{}
\begin{document}
\maketitle
\begin{abstract}
We show the following two extensions of the standard positive mass
theorem (one for either sign) : 
Let $({\cal N},g)$  and $({\cal N},g')$ be asymptotically flat Riemannian
3-manifolds with compact interior and finite mass, such that $g$ and
$g'$ are $C^{2,\alpha}$ and related via the conformal rescaling $g' = \phi^4 g$
with a $C^{2,\alpha}-$ function $\phi > 0$. Assume further that
the corresponding Ricci scalars satisfy $R \pm \phi^4 R'\ge 0$.
Then the corresponding masses satisfy  $m \pm m' \ge 0$. Moreover, in
the case of the minus signs, equality holds iff $g$ and $g'$ are
isometric, whereas for the plus signs equality holds iff both $({\cal N},g)$ and
$({\cal N},g')$ are flat Euclidean spaces. 

While the proof of the case with the minus signs is rather obvious, the
case with the plus signs requires a subtle extension of Witten's proof of
the standard positive mass theorem. The idea for this extension is due
to Masood-ul-Alam who, in the course of an application, proved
the rigidity part $m+m' = 0$ of this theorem, for a special conformal
factor. We observe that Masood-ul-Alam's  method extends to the general situation.

\end{abstract}

\newpage

The positive mass theorem of Schoen and Yau \cite{SY} and Witten \cite{EW}
is a mathematical result with a direct physical interpretation: It showed that the concept of
mass in relativity as defined in \cite{ADM} is useful. Moreover, it has also proved to be
an important tool in obtaining mathematical results of a more general nature within and
beyond Relativity. 
In such applications the positive mass theorem has  normally been used in combination
with a suitable conformal rescaling of the metric, and it is in one way or the other
important to keep control over the the mass in this process.
This applies to the Yamabe problem \cite{LP}, to Herzlich's proof of a Penrose-type 
inequality \cite{HERZ} and in particular to the uniqueness result for
non-degenerate static black holes by Bunting and Masood-ul-Alam \cite{BM}.

In these contexts the following two results (one for either sign) might be of interest.
A special case has already been proven and applied before, as will be outlined below.\\ \\
{\it Theorem.} Let $({\cal N},g)$  and $({\cal N},g')$ be asymptotically flat Riemannian
3-manifolds with compact interior and finite mass, such that $g$ and
$g'$ are $C^{2,\alpha}$ and related via the conformal rescaling $g' = \phi^4 g$
with a $C^{2,\alpha}-$ function $\phi > 0$. Assume further that
the corresponding Ricci scalars satisfy $R \pm \phi^4 R'\ge 0$.

Then the corresponding masses satisfy  $m \pm m' \ge 0$. Moreover, in
the case of the minus sign, equality holds iff $g$ and $g'$ are
isometric, whereas for the plus sign equality holds iff  both $({\cal N},g)$
and $({\cal N},g')$ are flat Euclidean spaces. \\

Due to their formal similarity  we could not resist presenting these two results in a unified
manner. However, their proofs as well as their interpretations and applications are quite
different as far as presently known. We first discuss these interpretations and applications
and postpone the technical part.

The bound on the mass given by the "-" part of the theorem has the following direct
interpretation. Note that in Newtonian theory it is clear that the mass of a system 
exceeds the mass of another one if the density of matter of the first
system exceeds the density of matter of the second system everywhere. In relativity one
cannot expect such a subadditivity property to hold in general because the gravitational
field also carries energy.  
Nevertheless, as matter density is represented on a time-symmetric slice by the
Ricci scalar, the "-" part of the above theorem is a result of this kind.
It may be interpreted by saying that, under conformal rescalings of time symmetric
data, the change of their "matter component" always dominates the change in their
"radiative component".

As to the "+" case of the theorem, it will clearly be meaningful only if
non-positive masses are allowed, at least a priori. Then the result may be interpreted
as above, and it might be relevant for quantum gravity.
On the other hand, the rigidity part of this case has proved particularly
interesting as a technical tool in uniqueness proofs for black holes,
which we recall here.

Generalizing the classical result due to Israel \cite{WI}, Bunting and
Masood-ul-Alam proved that the Schwarzschild black hole solution is the unique static,
asymptotically flat and appropriately regular vacuum spacetime with a
non-degenerate Killing horizon \cite{BM}. In essence their method
consists of performing, on the induced metric on the $t = const.$ slice, a suitable conformal
transformation which removes the mass, followed by applying the rigidity case of the
standard positive mass theorem. This result generalizes easily (namely by applying
formally the same conformal rescaling as in the vacuum case) to show the absence of regular
single scalar fields \cite{TZ} or the absence of rather special $\sigma-$model fields
\cite{HEUS} in spacetimes with non-degenerate horizons. With some effort a suitable
conformal rescaling could also be obtained in the Einstein-Maxwell case,
which yields uniqueness of the non-extreme Reissner-Nordstr\"om solution
\cite{AM1,PR,WS}. Furthermore, extensions of the vacuum and the electrostatic
case with include extreme horizons have also been obtained \cite{PC1,PC2}.
However, already in the coupled Einstein-Maxwell-dilaton case
apparently natural conformal rescalings do not produce (manifestly) non-negative
Ricci scalars as required for applying the standard version of the positive
mass theorem.

This motivated  Masood-ul-Alam \cite{AM2} to prove the rigidity case of the "+"- part of
the theorem above, for a particular conformal factor. He could then
apply this result to show, in the Einstein-Maxwell dilaton case without
magnetic fields and with non-degenerate horizons, the uniqueness of a 2-parameter family of
solutions found by Gibbons \cite{GG}. 

The  "+"-part of the theorem as formulated in this paper is useful for obtaining uniqueness 
proofs in more general situations, in particular for black holes in the presence of more 
general matter fields. Such results will be described elsewhere.

We finally remark here that it would also be desirable to obtain a "spacetime"-version
of our result. This means that we expect to obtain corresponding bounds on the
ADM 4-momentum \cite{ADM} by imposing suitable requirements on the gravitational Cauchy
data $(g,p)$ and on suitably conformally  rescaled data $(g',p')$. General results on the
conformal behaviour of such data \cite{JY}, and the "spacetime" formulation of the positive mass
theorem as given in Witten's original paper \cite{EW} suggest that this might be possible.\\

We consider the following class of manifolds.\\ \\
{\it Definition.} A Riemannian 3-manifold $({\cal N},g)$ is said to
be asymptotically flat with compact interior and to satisfy the mass 
decay conditions (AFCIMD) if ${\cal N}$ is the union of a
compact set ${\cal K}$ and an end ${\cal E}$  of topology $R^3$  minus a ball, and if
(with respect to some asymptotic structure which we suppress in our notation)
\begin{eqnarray}
\label{af1}
g - \delta & \in &  C^{2,\alpha}({\cal N}) \cap C^{2,\alpha}_{-\tau}({\cal E})  
\qquad \mbox{for}\,\, 0 < \alpha < 1, \, \frac{1}{2} < \tau < 1, \, \mbox{and}\\
\label{af2}
R & \in & L^{1}({\cal N}).
\end{eqnarray}
Here $\delta$ is the Kronecker symbol and $C^{2,\alpha}$ and $C^{2,\alpha}_{-\tau}$
denote H\"older spaces and weighted H\"older spaces respectively. For the
latter we adopt the weight index convention of Bartnik \cite{RB} (also chosen by
Lee and Parker \cite{LP}) which gives directly the growth at infinity, i.e.
 $f \in C^{2,\alpha}_{-\tau}$ for some function $f$ implies $f =
O(r^{-\tau})$ (and corresponding falloff conditions on the derivatives).

Some remarks on this AFCIMD definition are in order. 
The requirement $\tau < 1$ is not a restriction here but just
introduces a notation suitable to formulate the lemma below. Note in 
particular that (\ref{af1}) does allow  $g - \delta$ to fall off like $O(r^{-1})$.
Thus (apart from this subtlety) our definition  agrees with that required for
a Witten-type proof in (the appendix of) \cite{LP}. The name "mass decay condition" is
adopted from  Bartnik (c.f. Def. 2.1 and Sect. 4 of \cite{RB}) who requires, however,
weaker decay conditions formulated in terms of Sobolev spaces. 
The reason for formulating the present work in terms of H\"older spaces
is again the lemma on the uniqueness of the conformal structure given below,
which then becomes a rather obvious consequence of a known result.

Both within the H\"older as well as within the Sobolev setting, the AFCIMD conditions 
are the weakest ones which guarantee that the ADM mass
\begin{equation}
\label{adm}
m = \frac{1}{16\pi}\int_{S_{\infty}}(\partial_{j}g_{ij} - \partial_{i}g_{jj})dS^{i}
\end{equation}
is well defined and finite \cite{RB,PC3}. (Here and below, $dS^{i}$ denotes
the outward normal surface element to $S_{\infty}$, the sphere at infinity, and
repeated indices are summed over).

For what follows it is useful to recall the behaviour of the Ricci scalar under
conformal rescalings $g' = \phi^4 g$, viz.
\begin{equation}
\label{Rico}
\triangle \phi = \frac{1}{8}(R - \phi^4 R')\phi.
\end{equation}

We have the following lemma on uniqueness of the conformal structure. \\ \\
{\it Lemma.} Let $({\cal N},g)$ be a Riemannian manifold which 
satisfies the AFCIMD conditions as formulated in the definition above.
Then the same applies to $({\cal N}, g')$ with $g' = \phi^4 g$  iff
\begin{eqnarray}
\label{om}
\phi - 1 & \in & C^{2,\alpha}({\cal N}) \cap C^{2,\alpha}_{-\tau}({\cal E}) 
\qquad \mbox{for}\,\, 0 < \alpha < 1, \, \frac{1}{2} < \tau < 1, \,
\mbox{and}\\
\label{lom}
\triangle \phi & \in & L^{1}.
\end{eqnarray}
{\it Proof.} The result that $({\cal N}, g')$ is AFCIMD follows trivially
from (\ref{om}) and (\ref{lom}), using (\ref{Rico}).

On the other hand, requiring that $({\cal N}, g')$ is AFCIMD, (\ref{om}) is a
consequence of Theorem 2.4 of \cite{PC4}. Then (\ref{lom}) is obvious
from (\ref{af2}) and (\ref{Rico}).\hfill $\Box$ \\ 

A well known (or from (\ref{adm}) and (\ref{om}) easily proven) fact is that the
masses of two AFCIMD metrics $g$ and $g'= \phi^4 g$ are related by
\begin{equation}
\label{mpr}
m - m' = \frac{1}{2\pi} \int_{S_{\infty}} \nabla_{i}\phi~dS^{i}.
\end{equation}\\
{\it Proof of the theorem, part "-"}.
Applying Gauss' law to (\ref{Rico}) and using (\ref{om}) we can write
(\ref{mpr}) as
\begin{equation}
\label{mdif}
m - m' = \frac{1}{16\pi}\int_{\cal N}(R - \phi^4 R')\phi~dV
\end{equation}
where $dV$ is the volume element on $({\cal N},g)$. 
The inequality $m - m' \ge 0$ is then obvious from the assumption that 
$R - \phi^4 R'\ge 0$. Requiring now that $m = m'$, eqn. (\ref{mdif}) and 
the assumption on the Ricci scalars imply $R = \phi^4 R'$ on ${\cal N}$.
Thus, from (\ref{Rico}) we have  $\triangle \phi = 0$ on ${\cal N}$. 
But since $\phi \in C^{2,\alpha}$ and goes to 1 at infinity, this is only 
possible if $\phi = 1$ on ${\cal N}$, which had to be shown. 
\hspace*{1cm}\hfill $\Box$\\

The "+" part of the theorem will now be shown via Witten's techniques, as an
extension of the proof of Masood-ul-Alam \cite{AM1}. 

We consider the bundle of Dirac spinors $\Gamma$ as recalled, e.g. in
\cite{RB,LP,PT}, denote by $C^{2,\alpha}(\Gamma)$ and by $C^{2,\alpha}_{-\tau}(\Gamma)$ 
the H\"older spaces and weighted H\"older spaces of sections of $\Gamma$, respectively, 
and by ${\cal D}$ the Dirac operator. We adopt the notation 
 $\sigma_{ij} = \frac{1}{2}[e_i, e_j] = e_i e_j + \delta_{ij}$ for the Clifford
algebra with basis $e_{i}$.

To facilitate understanding of the following manipulations we recall
the so-called Lichnerowicz identity for $\Psi \in C^{2,\alpha}(\Gamma)$
(which is in fact due to Schr\"odinger, formula (74) of \cite{ES}),
\begin{equation}
\label{LW}
{\cal D}^2 \Psi = \nabla \ast \nabla  \Psi  +  \frac{1}{4} R  \Psi.  
\end{equation}
where $\nabla \ast \nabla = - g^{ij}\nabla_{i}\nabla_{j} = - \nabla^{i}
\nabla_{i}$ is the covariant Laplacian on spinors.
This relation implies, for solutions of the Dirac equation ${\cal D}\Psi = 0$, 
\begin{equation}
\label{lappsi}
\triangle \vert \Psi \vert^2 = \frac{1}{2} R \vert \Psi \vert^2 + 2 \vert \nabla
\Psi \vert^{2}.
\end{equation}
Again for solutions of the Dirac equation we then find, using (\ref{Rico}) and  (\ref{lappsi}) 
in the final step,
\begin{eqnarray}
\label{MS1}
\lefteqn{\nabla_{i}[\nabla^{i}|\Psi|^2 - 2 \phi^{-1}(\nabla_{i}\phi)| \Psi|^2 ] = {}} \nonumber \\
& & {}=  \triangle|\Psi|^{2} + 2\phi^{-2} (\nabla_{i}\phi) (\nabla^{i}\phi) |\Psi|^2 
- 2 \phi^{-1} (\triangle \phi ) |\Psi|^{2} - 2 \phi^{-1}(\nabla^{i}\phi) \nabla_{i}|\Psi|^{2}
  = {} \nonumber\\
& & {} = \frac{1}{4} (R  + \phi^4 R') \vert \Psi \vert^2 + 
2 \vert \nabla_{i} \Psi - \phi^{-1}(\nabla_{i} \phi) \Psi \vert^2.
\end{eqnarray}
{\it Proof of the theorem, part "+"}.
In analogy with \cite{RB} and \cite{AM1}, we show first that the Dirac operator
\begin{equation}
\label{DH}
{\cal D} : C^{2, \alpha}_{-\tau}(\Gamma) \rightarrow C^{1, \alpha}_{-\tau-1}(\Gamma)
\qquad \mbox{for}\,\, 0 < \alpha < 1, \, 0 < \tau < 2
\end{equation}
is an isomorphism. Passing to Sobolev spaces $W^{1,q}_{-\epsilon}$
(as defined e.g. in \cite{RB}) via the embedding 
$C^{k, \alpha}_{-\tau}(\Gamma) \subset W^{k,q}_{-\epsilon}(\Gamma)$ for any
$ k\ge 0,\, q > 1,\,\epsilon < \tau$, standard results (\cite{MC,RL}) imply
that, for 
$q > 1, \, 0 < \epsilon < 2$,
\begin{equation}
\label{DS}
{\cal D} : W^{2, q}_{-\epsilon}(\Gamma)  \rightarrow  W^{1, q}_{-\epsilon - 1}(\Gamma)
\end{equation}
is Fredholm with adjoint
\begin{equation}
{\cal D} = {\cal D}^{*} : W^{2, q'}_{\epsilon + 1 - n}(\Gamma)  \rightarrow 
W^{1, q'}_{\epsilon - n}(\Gamma)
\end{equation}
for $q' = (1 - q^{-1})^{-1}.$
If $\Psi \in \mbox{ker}~{\cal D}$, then $|\Psi| \rightarrow 0$ at infinity. The strong maximum 
principle applied to relation (\ref{MS1}) then shows that $ |\Psi|^{2} = 0$. Hence (\ref{DS}) 
and its adjoint have trivial kernels and are isomorphisms. The regularity claimed in (\ref{DH}) 
then follows from the ellipticity of the Dirac operator \cite{CM}.

Let now $\Psi_{0}$ be a spinor which is constant at infinity. Then there is a spinor
$\Psi$ such that
\begin{eqnarray}
{\cal D}\Psi & = & 0, \\
\Psi - \Psi_{0} & \in & C^{2,\alpha}_{-\tau}(\Gamma).
\end{eqnarray}

We can thus find  a solution $\Psi$ of the Dirac equation which tends to a
prescribed constant spinor. We now integrate (\ref{MS1}) over a ball with boundary 
$S_{\rho}$ (a sphere of coordinate radius $\rho$), use the requirement $R + \phi^4 R' \ge 0$ 
of the theorem and apply Gauss' law to obtain
\begin{eqnarray}
0 & \le & \int_{S_{\rho}} [\nabla_{i} \vert \Psi \vert^2 -  
2 |\Psi|^{2} \nabla_{i}\,\mbox{ln} \phi] dS^{i} = {} \nonumber \\
& & {} =  2 \int_{S_{\rho}} (\langle \Psi, \nabla_{i} \Psi \rangle -
\vert \Psi \vert^2 \nabla_{i}\,\mbox{ln}\phi) dS^{i}= {} \nonumber \\
\label{MS2}
& & {} = 2 \int_{S_{\rho}} (\langle \Psi, \sigma_{ij}
\cdot \nabla_j \Psi \rangle - \vert \Psi \vert^2 \nabla_{i}\,
\mbox{ln}\phi)dS^{i}.
\end{eqnarray}
Finally, passing to the limit $\rho \rightarrow \infty$, the first surface integral 
in $(\ref{MS2})$ is known to give the mass \cite{LP}, whereas the second one is evaluated 
by virtue of (\ref{om}) and (\ref{mpr}). This yields
\begin{equation}
\label{MS3}
0 \le 8\pi \vert\Psi_{0} \vert^{2} m -
4\pi \vert\Psi_{0} \vert^{2} (m - m') = 
4\pi \vert\Psi_{0} \vert^{2} (m + m').
\end{equation}

To show the rigidity case we note that using $m + m' = 0$ in the integral of (\ref{MS1}) 
yields $\nabla( \phi^{-1}\Psi) = 0$, $R = 0$ and $R' = 0$. The existence of a covariantly
constant spinor implies by a standard argument (see e.g. \cite{RB}) that $({\cal N},g)$ is
flat, so in particular $m = 0$. Therefore we also have $m' = 0$, and applying the standard 
positive mass theorem on $({\cal N},g')$ finishes the proof. \hfill $\Box$ \\ \\
{\it Acknowledgement.} I am grateful to Piotr Chru\'sciel for helpful
discussions and for useful comments on the manuscript, and to the referee
for suggesting improvements.

\end{document}